\documentclass[epj,twocolumn]{webofc}
\usepackage[varg]{txfonts}  
\usepackage{amsmath,amssymb}
\usepackage{multirow}
\graphicspath{ {./figs/} }
\newcommand{\PR}{{ Phys. Rev. }}
\newcommand{\PRL}{{ Phys. Rev. Lett. }}

\newcommand{\etal}{{\em et~al.}}
\woctitle{Advances in Dark Matter and Particle Physics}
\begin{document}
\title{Study of the one-way speed of light anisotropy with particle beams}
\author{\firstname{Bogdan} \lastname{Wojtsekhowski}\inst{1}\fnsep\thanks{\email{bogdanw@jlab.org}}}

\institute{Thomas Jefferson National Accelerator Facility, Newport News, Virginia 23606 USA}

\abstract{Concepts of high precision studies of the one-way speed of light anisotropy are discussed.
The high energy particle beam allows measurement of a one-way speed of light anisotropy (SOLA) via analysis 
of the beam momentum variation with sidereal phase without the use of synchronized clocks.
High precision beam position monitors could provide accurate monitoring of the beam orbit
and determination of the particle beam momentum with relative accuracy on the level of $10^{-10}$,
which corresponds to a limit on SOLA of $10^{-18}$ with existing storage rings.
A few additional versions of the experiment are also presented.
}
\maketitle
\section{Introduction}
A constant and isotropic speed of light is a key postulate of the special relativity theory, SR,
as formulated by A.~Einstein in 1905~\cite{AE1905}.
In a search for a light-related medium, ether, A.~Michelson~\cite{AM1881} developed an extremely sensitive method 
of measurement of the anisotropy of the two-way speed of light,  $c_2$, which is an average of the speeds 
in two opposite directions.
A modern experiment of this type~\cite{MN2015} reached an extremely high precision, 
$10^{13}$ times higher than was obtained in the Michelson-Morley experiment~\cite{MM1887}, 
and found that $c_2$ is isotropic at least to the level of $10^{-18}$.
The one-way speed of light anisotropy tells us about the directional variation of 
a difference in the speed of light in two opposite directions.
Current studies of SOLA are motivated by predictions in string and quantum gravity theories~\cite{DM2005}, 
as well as by general interest in physics beyond the Standard Model.
The one-way speed of light, $c_1$, could be measured by using two synchronized clocks.
However, there is a very significant difficulty in realizing clock synchronization at the required precision.
If observed, variation of the $c_1$ speed of light could be an indication of a quantum gravity effect in the laboratory. 

Several tests have been proposed for the study of $c_1$ which do not require clocks because they address 
the difference between the speed of light at different polarizations or the difference between 
the speed of light and the speed of electrons in the same direction.
The latter could be explored using threshold reactions such as vacuum Cherenkov 
radiation and photon decay to an electron-positron pair~\cite{MH2009}.
The maximum attainable speed of electrons could also have directional anisotropy.
A Compton process was used recently in an experiment~\cite{JB2010}, which provided a constraint 
on a sidereal time variation of a combination of the maximum attainable speed of electrons and the one-way speed of light.
Many (but not all) of the SOLA experiments were analyzed in terms of mSME developed 
by A.~Kostelecky and collaborators~\cite{AK2011}.

\section{Concept of the method}
The trajectory of the high energy beam in a transverse magnetic field provides an exciting opportunity 
to investigate SOLA~\cite{BW2014}.  
Indeed, according to SR, the momentum of the particle is defined by the difference between its speed and 
a maximum attainable speed for the particle (assumed from now on to be equal to the speed of light).
The relativistic expression for the particle momentum, $p = mv/\sqrt{(c-v)\cdot(c+v)}$,
shows that when $v$ approaches $c$, the $p$ value is very sensitive to the variation of $(c-v)$.
For a charged particle moving in a transverse magnetic field, according to the conventional 
Lorentz force, we should have a constant absolute value of the velocity vector $\vec{v}$.
If the speed of light, $c_1$, depends on direction, the experimental observable in the laboratory, 
a particle momentum, will also change with sidereal time because of rotation of the laboratory
together with the Earth, and its relative variation would be enhanced by a factor of $\gamma^2$.

Many modern accelerators are equipped with beam position monitors which have an accuracy 
of a few microns or better in a single measurement.
Using a typical magnetic system with a dispersion function of a few meters, it should be possible
to measure beam momentum to a few $10^{-6}$ precision many times per second.
There are several accelerators with a beam gamma factor of $10^4$, which provides a
boost factor of $10^8$ in sensitivity for the speed of light variation. 
Such a boost could allow us reach the quantum gravity domain whose possible onset is $10^{-17}$~\cite{DM2005}.

In addition to a sidereal time variation, very short-term changes are also well motivated physics, e.g.
transient-in-time effects~\cite{MP2013},
and they could be detected via particle momentum changes in a high energy accelerator.
Correlated measurements at several facilities could provide a realistic way to search for 
such a phenomenon~\cite{PT2015}.

\section{Beam experiment at CEBAF}
The CEBAF accelerator has ten 180$^\circ$ magnetic arcs with beam energies varying from one to eleven GeV.
There are no accelerating cavities in the arcs, and in each arc the beam energy is constant 
(after correction for a small and stable synchrotron radiation effect).
Formulated in the Ref.~\cite{BW2014} plan is a measurement of the momentum of the beam at two points 
at the beginning and the end of each arc by using beam deflection in arc magnets and beam position monitors.
The analysis of the beam momenta ratio at these two points as a function of time would be
independent of beam energy variation, which is typically of $10^{-4}$.

The data for such an experiment would be collected during normal CEBAF operation for hadron
physics experiments~\cite{CEBAF_2015}.
Similar measurements could be performed at the KEK high energy electron/positron rings and
at LHC for protons, as well as at a number of X-ray facilities.

The main limitation in such an experiment is due to instabilities in the magnetic 
system and beam line geometry due to a number of effects such as ground movement (a tidal effect), 
oscillation in power supplies for magnetic elements, daily variation of the tunnel temperature and many others.
However, analysis of the data over many days should allow effective selection of the signal
with sidereal periodicity.

\section{Beam experiments at positron/electron rings}
There are storage rings where one can take advantage of closely located orbits of counter-propagating
beams and analyze the ratio of the beam momenta in which the instabilities of the magnetic system and geometry 
drifts are canceled out because they have identical effects on both beams~\cite{BW2015}.

Two electron/positron storage rings, CESR and VEPP-4, currently operate with a common magnetic system 
for both beams and have large beam gamma factors. 
In these storage rings, the electron-to-positron orbit differences, originating mainly from beam 
energy loss due to synchrotron radiation and coherent effects, are less than 0.5 mm.
Such a configuration of the experiment is able to probe the CPT-even terms of the Lorentz violation with
precision limited only by the short term stability of the beam position monitors (assuming
that the same BPMs and electronics are used from both beams).

Analysis of the first CESR data allows us to put a limit on $\delta c_1/c_1 \leqslant 10^{-14}$.
Currently, experiments are active in both accelerators~\cite{DR2016,SN2016}.

\section{Experiments at the X-ray synchrotron radiation storage rings}
There are several storage rings with multi GeV electron beams which are used 
as X-ray sources to do photon physics~\cite{SPRING-8,APS,ESRF}.
Each of them provides dozens of X-ray beam lines with very stable direction 
(on the order of 0.1 micro radian) and position (on the order of a micro meter) of photons
at users' instruments.

The beam orbit in a typical X-ray storage ring is subject to regular corrections which
allow the X-ray beam to be kept stable.
These corrections complicate the use of the beam orbit data for the speed of light investigation.
However, stabilization of the X-ray beam directions also leads to stabilization of the electron
beam directions at the location of X-ray emission in the insertion devices.

As a result, the magnetic field integral on the beam orbit between two X-ray beam emission points
would provide a perfect measure of the beam momentum and open the possibility
of SOLA investigation at many synchrotron radiation facilities.

\section{Experiment with a muon storage ring}
A high precision measurement of the muon anomalous magnetic moment, the muon $(g-2)$ experiment, 
was performed at a specialized storage ring with decay in flight of 3.09 GeV/c muons~\cite{muon_g-2}.
It was done via analysis of the muon decay rate observed in the detectors with the 5-parameter fit:
\mbox{$N(t) \,=\, N_0 \, e^{-(t/{\gamma \tau})} \left[1 \,+\, A\cos{(\omega_a t \,+\, \phi)}\right]$}.
The data were also used to evaluate possible Lorentz invariance violations in a muon sector of mSME~\cite{muon_LIV}.
In that analysis, the spin precession frequencies $\omega_a$ for positive and negative 
muons were analyzed for potential sidereal time variation. 

The muon spin rotation frequency at a given magnetic field value is defined by the muon magnetic moment and its momentum.
The relative angle between the momentum and spin directions is defined by the anomalous magnetic moment 
and momentum rotation angle.
The change in the observed muon decay rate between detectors could be used as a measure of the momentum rotation angle advance.
As the speed of light variation induces the particle momentum variation, the local spin rotation frequency 
also varies, as does the direction of the muon spin at each detector, $D$.
The SOLA effect could be searched for by taking into account the detector location in contrast to 
the performed analysis~\cite{muon_LIV}, where the frequency $\omega_a$ was averaged over all detectors 
along the orbit before a search for a sidereal time variation.

We propose that in the fit function above, the $\phi$ should be replaced by:
$\phi + \omega_a \int \frac {\Delta p}{p} dt \approx \phi_0 + \frac{\omega_a}{ \omega_{rev} } \, \gamma^2 {a_\perp} \times$ \\
\mbox{$\times [ \cos(\Omega_{_E} t + \psi) \cos \phi{_{_D} } + \sin (\Omega_{_E} t + \psi) \sin \phi_{_D} \sin \theta_{ring} ] $, } 
where $\gamma$ is the muon beam Lorentz factor, $a_\perp$ is the SOLA component in the direction 
perpendicular to the Earth's axis of rotation, $\omega_{rev}$ is the beam revolution frequency in the ring,
$\Omega_{_E}$ is the Earth's rotation frequency, $\psi$ is a SOLA phase, 
$\phi_{_D}$ is the location of the individual detector along the beam orbit, and $\theta_{ring}$ is the laboratory (ring) latitude.
The muon life time term $\gamma \tau$ also needs a similar adjustment.
The proposed analysis will provide a limit on the parameter $a_\perp$ and, as a result, a first estimate for
a sidereal time variation of the maximum attainable speed of a muon.

\section{Analysis of a resonance production rate}
The asymmetric electron-positron collider allows an interesting opportunity to monitor
the speed of light by observing the production rate variation when the combined beam
energy is set slightly off resonance~\cite{BW2014}.
In the case of $\Upsilon$ resonance, the observed relative width is $\Gamma/\mathrm{M} \sim 10^{-3}$.
Indeed, due to a large difference in gamma factors, the speed of light variation induces changes 
in the momentum for a high energy beam $(h.e.b.)$ much larger than those for a lower energy beam.

As a result, the total energy in the lab system varies with sidereal time, and the resonance production rate also varies.
The size of the rate variation could be estimated 
as \mbox{$\frac{\delta c_1}{c_1} \times \gamma_{h.e.b.}^2 \times \mathrm{M}/\Gamma$}.
At the projected luminosity of the Belle-2 experiment~\cite{Belle-2}, statistics collected every day would be
of $10^8 \,\, \Upsilon$ events, which would correspond to a sensitivity 
for $\delta c_1/c_1$ on the level of $10^{-15}$.

\section{Polarized electron-positron beams at a storage ring}

The measurement of the electron and positron anomalous magnetic moments ratio, 
$\mu^\prime_+/\mu^\prime_-$, at VEPP-2m~\cite{IV1987} found that precession frequencies are equal 
to each other at the relative level $1.2 \cdot 10^{-7}$ (a record precision for the time of that experiment).
The current best limit for $(g_+ - /g_-)/g_{aver.}$ of $2 \cdot 10^{-12}$ was obtained using the particle trap method~\cite{PDG2016}. 

As a result, we can consider~\cite{BW2014} the data from the storage ring experiment as a comparison of 
the electron and positron beam momenta when the beams are moving in opposite directions
in a transverse magnetic field and make an estimate of the achievable sensitivity  
for the $c_1$ variation on the level of $1 \cdot 10^{-13}$ from those 1987 data.
Obviously, this could be improved by two to three orders of magnitude using a higher 
beam energy, e.g. at VEPP-4, and advanced spin resonance detection methods~\cite{YS2016}.

\section*{Acknowledgments}

The author would like to thank the organizing committee of the ``Advances in Dark Matter and Particle Physics" 
workshop for support and the invitation to give a talk. 
I greatly appreciate comments from D.~Budker, Yu.~Shatunov and V.~Zelevinsky, and collaborations 
with W.~Bergan, G.~Franklin, V.~Lebedev, S.~Nikitin, G.~Nowak, Y.~Roblin, D.~Rubin, B.~Schmookler, and B.~Vlahovic.
This work was supported in part by the NCCU NSF CREST Center, and by Department of Energy (DOE) 
contract number DE-AC05-06OR23177, under which the Jefferson Science Associates operates 
the Thomas Jefferson National Accelerator Facility.


\begin{thebibliography}{0}
%
\bibitem{AE1905}  A.~Einstein, Annalen der Physik, {\bf 17}, 891 (1905).
%
\bibitem{AM1881} A.A. Michelson, American Journal of Science, {\bf 22}, 120 (1881).
%
\bibitem{MN2015} M.~Nagel \etal, Nature Communications,  {\bf 6}, 8174 (2015).
%
\bibitem{MM1887} A.A.~Michelson and E.W.~Morley, Am. J. Sci., {\bf 34}, 333 (1887).
%
\bibitem{DM2005} D.~Mattingly, Living Rev. Rel. {\bf 8}, 5 (2005).
%
\bibitem{MH2009} M.A.~Hohensee \etal, \PRL {\bf 102}, 170402 (2009).
%
\bibitem{JB2010} J.-P.~Bocquet \etal, \PRL {\bf 104}, 241601 (2010).
%
\bibitem{AK2011} V.A.~Kostelecky and N.~Russell, Rev. Mod. Phys. {\bf 83}, 11 (2011).
%
\bibitem{BW2014} B.~Wojtsekhowski, Euro Physics Letters, {\bf 108}, 31001 (2014); arXiv:1409.6373.
%
\bibitem{MP2013} M.~Pospelov \etal, \PRL {\bf 110}, 021803 (2013).
%
\bibitem{PT2015} D.~Budker and A.~Derevianko, Physics Today 68(9), 10 (2015).
%
\bibitem{CEBAF_2015} B.~Wojtsekhowski and Y.~Roblin, proposal PR12-15-002, JLab PAC43, 2015.
%
\bibitem{BW2015} B.~Wojtsekhowski, arXiv:1509.02754.
%
\bibitem{DR2016} D.~Rubin \etal, private communication, Aug. 2016.
%
\bibitem{SN2016} S.~Nikitin \etal, private communication, Oct. 2016.
%
\bibitem{SPRING-8} \url{http://www.spring8.or.jp/en/}
%
\bibitem{APS} \url{https://www1.aps.anl.gov}
%
\bibitem{ESRF} \url{http://www.lightsources.org/facility/esrf}
%
\bibitem{muon_g-2} G.W.~Bennett \etal, \PR D {\bf 73}, 072003 (2006).
%
\bibitem{muon_LIV} G.W.~Bennett \etal, \PRL {\bf 100}, 091602 (2008).
%
\bibitem{Belle-2} T.E.~Browder \etal, arXiv:0802.3201.
%
\bibitem{IV1987} I.B.~Vasserman \etal, Phys. Lett. B, {\bf 198} 302 (1987).
%
\bibitem{PDG2016} C.~Patrignani \etal. (Particle Data Group), Chin. Phys. C, 40, 100001 (2016).
%
\bibitem{YS2016} Yu.M.~Shatunov, private communication, Sept. 2016.
%
\end{thebibliography}
\end{document}